# Absorption of Transverse Spin Current in Ferromagnetic NiCu: Dominance of Bulk Dephasing over Spin-Flip Scattering

Youngmin Lim[1], Shuang Wu[1], David A. Smith[1], Christoph Klewe[2], Padraic Shafer[2], Satoru Emori[1,*]

1. Department of Physics, Virginia Tech, Blacksburg, Virginia 24060

2. Advanced Light Source, Lawrence Berkeley National Laboratory, Berkeley, California 94720

*email: semori@vt.edu

**In ferromagnetic metals, transverse spin currents are thought to be absorbed via dephasing – i.e., destructive interference of spins precessing about the strong exchange field. Yet, due to the ultrashort coherence length of ≈1 nm in typical ferromagnetic thin films, it is difficult to distinguish dephasing in the bulk from spin-flip scattering at the interface. Here, to assess which mechanism dominates, we examine transverse spin-current absorption in ferromagnetic NiCu alloy films with reduced exchange fields. We observe that the coherence length increases with decreasing Curie temperature, as weaker dephasing in the film bulk slows down spin absorption. Moreover, nonmagnetic Cu impurities do not diminish the efficiency of spin-transfer torque from the absorbed spin current. Our findings affirm that transverse spin current is predominantly absorbed by dephasing inside the nanometer-thick ferromagnetic metals, even with high impurity contents.**

Spin currents underpin a variety of fundamental condensed-matter phenomena and technological applications [1–3], especially those based on magnetic materials. Of particular interest is coherent *transverse* spin current, where the flowing spins are uniformly polarized transverse to the magnetization. This spin current generates a spin-transfer torque that can switch a nanomagnetic memory or drive a GHz-range oscillator [4–6]. While spins may be carried by magnons [7] and phonons [8], they are often primarily carried by electrons in practical metallic multilayers incorporating ferromagnetic thin films. It is therefore crucial to understand the nanoscale transport of electron-mediated transverse spin current in ferromagnetic metals.

A spin current in any material ultimately becomes absorbed (loses coherence) within a finite length scale [1]. In ferromagnetic metals, transverse spin-current absorption can occur via *dephasing* [9–11], i.e., destructive interference of coherent spins that precess about the magnetic exchange field. The dephasing mechanism is illustrated in Fig. 1: The transverse electronic spins enter the ferromagnetic metal with a wide distribution of incident wavevectors; these spins traverse and precess about the magnetic exchange field at different rates, thereby averaging out the net transverse polarization (destroying the phase coherence) of the spin current within a finite length scale. Another possible mechanism of spin-current absorption is diffusive *spin-flip scattering* [12]. When electrons carrying the spin current are scattered, e.g., by impurities or an interface, the orientation of the propagating spins may be flipped to various orientations.

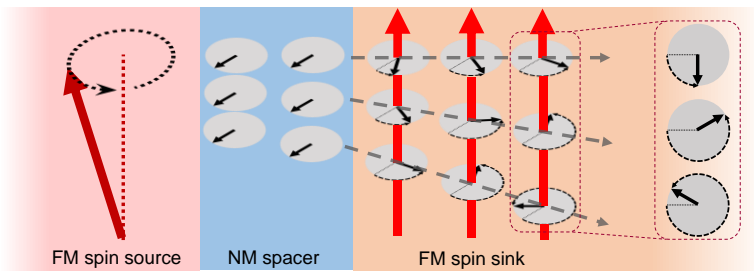

FIG. 1. Dephasing of a transverse spin current generated by FMR in the ferromagnetic (FM) spin source. The propagating spins are coherent in the normal metal (NM) spacer – as illustrated by the aligned black arrows – but they enter the spin sink with different incident wavevectors. In the FM spin sink, the spins precess about the ferromagnetic exchange field (red vertical arrows) by different amounts, thereby losing phase coherence.

Prior experiments [13] have quantified the absorption length scale – i.e., coherence length $\lambda_c$ – of transverse spin current through ferromagnetic resonance (FMR) spin pumping [14]. These experiments indicate $\lambda_c \approx 1$ nm from the ferromagnetic film thickness where the measured spin absorption saturates. This ultrashort $\lambda_c$ is presumably due to rapid dephasing [9–11] from the strong ferromagnetic exchange field of $\gg 100$ T [15]. Hence, the conventional wisdom is that transverse spin current is absorbed via dephasing, rather than spin-flip scattering. However, $\lambda_c \approx$ 1 nm corresponds to a nominal film thickness of a few lattice parameters, likely just at the threshold of forming a continuous film layer. Spin-flip scattering at the "interface" could be significant for such ultrathin ferromagnets. A plausible alternative explanation for $\lambda_c \approx 1$ nm is that interfacial spin-flip scattering saturates at the ferromagnetic thickness of ≈1 nm. Therefore, it remains a challenge to distinguish spin-flip scattering at the interface from spin dephasing in the ferromagnet's bulk.

In this Letter, we experimentally address the following fundamental question: Which mechanism – spin dephasing or spin-flip scattering – is responsible for the ultrashort coherence length $\lambda_c$ of transverse spin current in ferromagnetic metals? By employing the FMR spin pumping protocol similar to Ref. [13], we quantify $\lambda_c$ for ferromagnetic Ni films alloyed with nonmagnetic Cu that reduces the ferromagnetic exchange strength. Our hypothesis is that $\lambda_c$ must increase with increasing nonmagnetic Cu impurity content, if dephasing in the bulk is dominant. Alternatively, if spin-flip scattering at the interface is significant, $\lambda_c$ is expected to remain unchanged or perhaps decrease as the Cu impurities further enhance electronic scattering. Thus, testing the above hypothesis permits us to confirm – or refute – the long-held notion that dephasing in the ferromagnet's bulk drives transverse spin-current absorption. It is also timely to examine basic spin transport in NiCu alloys, which have attracted attention for their reportedly sizable spin-orbit effects [16–18] that may hold promise for spintronic devices.

Ni and Cu readily form homogeneous solid solutions, permitting continuous tuning of ferromagnetic exchange while maintaining the same face-centered cubic structure in NiCu alloys. Figure 2 summarizes the Curie temperatures $T_C$ – used as the metric for the ferromagnetic exchange strength [15] – of 10-nm-thick Ni, $Ni_{80}Cu_{20}$, and $Ni_{60}Cu_{40}$ films. We limit the maximum Cu content to 40 at.% to attain ferromagnetism close to room temperature, where our FMR spin pumping measurements were performed. Additional information on the growth and characterization of these films are included in the Supplementary Material. The measured values of $T_C$ are within ≈10% of prior literature reports for bulk Ni and NiCu [19,20]. The monotonic drop in $T_C$ seen in Fig. 2 verifies that the Cu impurities dilute the ferromagnetic exchange.

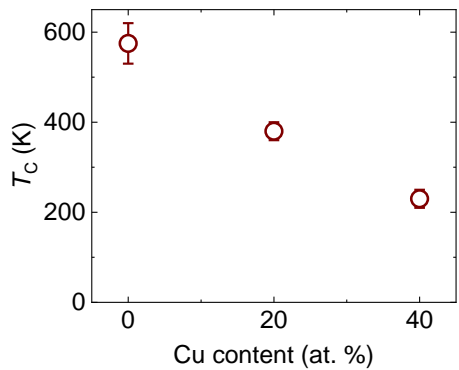

FIG. 2. Compositional dependence of the Curie temperature $T_C$ in 10-nm-thick NiCu films.

To derive $\lambda_c$, we conducted FMR spin pumping measurements on film stacks Si-SiO$_2$(substrate)/Ti(3)/Cu(3)/Ni$_{80}$Fe$_{20}$(10)/Ag(5)/Ni(Cu)(0-10)/Ti(3), where Ni(Cu) denotes the Ni, Ni$_{80}$Cu$_{20}$, or Ni$_{60}$Cu$_{40}$ "spin sink." The Ti/Cu seed bilayer promotes narrow FMR linewidths (minimizing two-magnon scattering [21]) in the NiFe "spin source," crucial for straightforward spin pumping measurements. The Ag spacer suppresses direct magnetic coupling between the NiFe source and Ni(Cu) sink, such that spin transport from the source to the sink is mediated solely by electrons without complications from magnon interactions [22]. Ag is selected as the spacer, instead of the oft-used Cu, to reduce atomic intermixing at the spacer/Ni(Cu) interface.

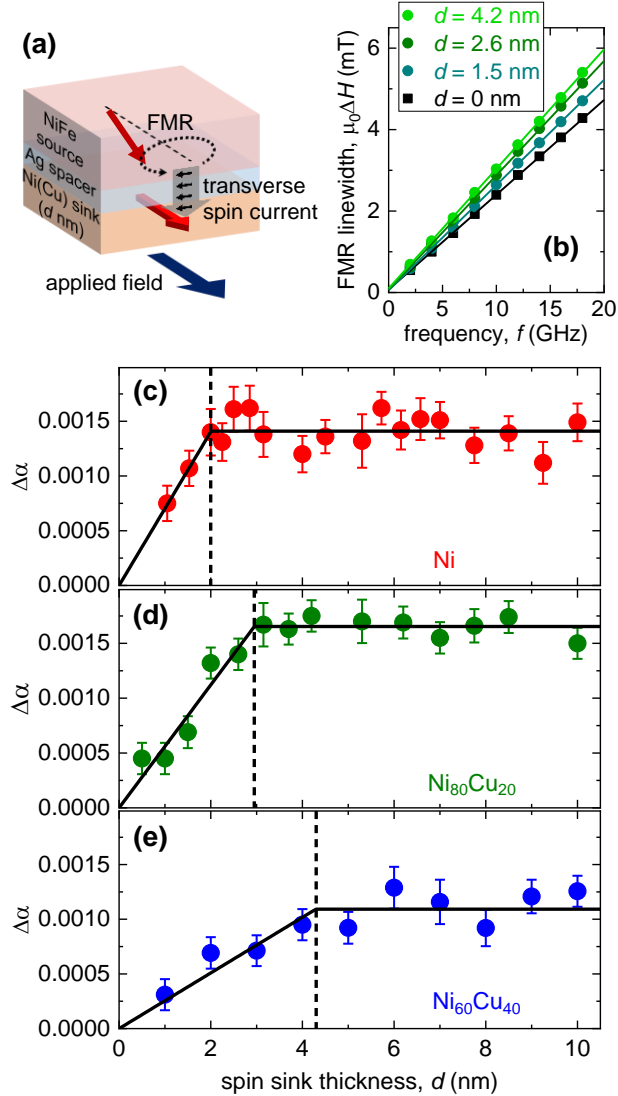

FIG. 3. (a) Illustration of FMR spin pumping with the NiFe spin source and the Ni(Cu) spin sink. (b) Frequency dependence of the FMR linewidth for different Ni$_{80}$Cu$_{20}$ spin sink thicknesses $d$. (c-e) Nonlocal damping enhancement $\Delta\alpha$ as a function of $d$, where the spin sink is (c) Ni, (d) Ni$_{80}$Cu$_{20}$, and (e) Ni$_{60}$Cu$_{40}$. The solid black lines indicate the fits with Eq. 1. The vertical dashed lines indicate the coherence length $\lambda_c$ extracted from the fits.

In the spin pumping scheme (Fig. 3(a)), a microwave field from a coplanar waveguide excites FMR in the NiFe source, such that the magnetization oscillates about the in-plane applied bias magnetic field. FMR generates a coherent ac spin current polarized transverse to the oscillation axis. This spin current is pumped through the nonmagnetic Ag spacer and into the Ni(Cu) sink. Since the thickness of Ag here is much smaller than the spin diffusion length of ~100 nm [12,23], the coherent spin current propagates with negligible absorption in the spacer [14,24]. The polarization of the spin current is transverse to the magnetization of the Ni(Cu) sink, which is set by the applied field. The FMR condition of the Ni(Cu) layer is sufficiently far from that of the NiFe source, so Ni(Cu) serves as a passive sink that receives the spin current from the NiFe source.

Any spin-current absorption in the Ni(Cu) sink constitutes an additional loss of spin angular momentum, which manifests in an enhancement of Gilbert damping $\Delta\alpha$ in the NiFe source [14,25]. As shown in Fig. 3(b), the total measured Gilbert damping parameter $\alpha$ is obtained from the linear slope of the FMR linewidth $\Delta H$ plotted against the microwave frequency $f$, $\mu_0 \Delta H = \mu_0 \Delta H_0 + \frac{2\pi}{\gamma}\alpha f$, where $\mu_0 \Delta H_0 < 0.1$ mT is the inhomogeneous linewidth broadening and $\frac{\gamma}{2\pi} = 29.8$ GHz/T is the gyromagnetic ratio for NiFe. By averaging samples from seven deposition runs, the baseline Gilbert damping parameter of NiFe/Ag without a Ni(Cu) sink is found to be $\alpha_0 = 0.00693 \pm 0.00014$, similar to other reports on NiFe thin films [26,27]. Figure 3(b) shows an increased slope of $\Delta H$ vs $f$ with finite Ni(Cu) sink thickness. This observation signifies a nonlocal damping contribution, $\Delta\alpha = \alpha - \alpha_0$, due to spin absorption in the sink. Figure 3(c-e) summarizes the dependence of spin absorption, captured by $\Delta\alpha$, on spin-sink thickness $d$. For each $d$, an averaged $\alpha$ was obtained by measuring at least three separate sample pieces. The error bars for $\Delta\alpha$ are primarily from the scatter in $\alpha_0$.

For each Ni(Cu) sink composition, $\Delta\alpha$ rises at small $d$ and then saturates (Fig. 3(c-e)). This behavior is consistent with spin-current absorption within a finite depth in the sink, such that there is essentially no additional absorption at $d \gtrsim \lambda_c$. We quantify $\lambda_c$ by fitting our experimental data of $\Delta\alpha$ vs $d$. One possible approach is to employ a modified drift-diffusion model [28–30], but this involves multiple free parameters (e.g., complex transmitted spin-mixing conductance [11,31]) that could produce overdetermined fits. Instead, we employ a simpler empirical fitting function employed by Bailey *et al.* [13,32,33] with only two parameters, i.e., $\lambda_c$ and $\Delta\alpha_{\text{sat}}$:

$$\Delta\alpha = \frac{\Delta\alpha_{\text{sat}}}{\lambda_c}\big(1 - H(d - \lambda_c)\big)d + \Delta\alpha_{\text{sat}} H(d - \lambda_c), \quad (1)$$

where $H(d - \lambda_c)$ is the Heaviside step function centered at $d = \lambda_c$. From the resulting fits in Fig. 3(c-e), we note that $\Delta\alpha_{\text{sat}}$ is slightly higher for the Ni$_{80}$Cu$_{20}$ sink whereas it is lower for Ni$_{60}$Cu$_{40}$. We attribute this variation in $\Delta\alpha_{\text{sat}}$ to the different spin-mixing conductances that depend on the effective spin susceptibilities in these magnetic spin sinks [34–37]. We emphasize, however, that our focus here is on the length scale of transverse spin-current absorption, $\lambda_c$.

The values of $\lambda_c$ from the fits with Eq. 1 are well over $\lambda_c = 1.2 \pm 0.1$ nm of $Ni_{80}Fe_{20}$ alloy from Ref. [13]. Specifically, we obtain $\lambda_c = 2.0 \pm 0.2$ nm for Ni, $3.0 \pm 0.2$ nm for $Ni_{80}Cu_{20}$, and $4.3 \pm 0.5$ nm for $Ni_{60}Cu_{40}$. These values exceed several atomic monolayers, strongly pointing to spin absorption in the *bulk* of the sink layer rather than at its interface.

The observation of longer $\lambda_c$ with increasing nonmagnetic Cu content already suggests the essential role of dephasing. To gain further insight into how $\lambda_c$ scales with the diluted ferromagnetic exchange (i.e., decreasing $T_C$), we plot $\lambda_c$ against the inverse of $T_C$ for the Ni(Cu) compositions investigated in our work, along with $Ni_{80}Fe_{20}$ from Ref. [13]. Figure 4 illustrates the central finding of this study: $\lambda_c$ scales inversely with the ferromagnetic exchange strength (represented by $T_C$). The consistent explanation is that decreasing exchange – hence weaker dephasing – from the nonmagnetic Cu impurities enables the transverse spin current to remain coherent over a distance well above ≈1 nm. While the Cu impurities enhance electronic momentum scattering (e.g., with the mean free path possibly reduced to ≈1 nm [38]), we do not see evidence for spin-flip scattering dominating over spin dephasing within the studied composition range of NiCu. Our finding indicates that in these Ni-based systems, spin dephasing in the bulk remains dominant over interfacial or impurity-induced spin-flip scattering.

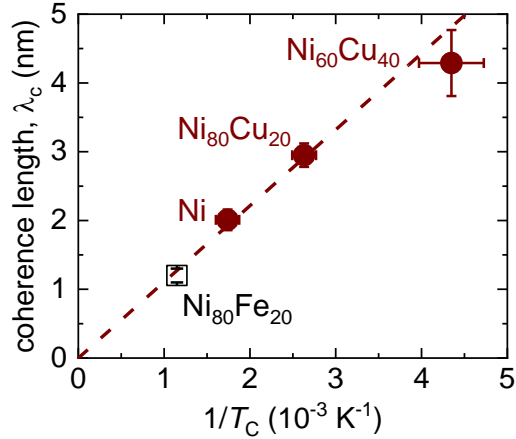

FIG. 4. Transverse spin-current coherence length $\lambda_c$ plotted against the inverse of the Curie temperature $T_C$. The data point for $Ni_{80}Fe_{20}$ is from Ref. [13].

The bulk nature of dephasing in these ferromagnets is distinct from prior reports on proximity-magnetized Pd and Pt films, in which the induced magnetic order is confined to a few monolayers at the interface [33,39,40]. It is also noteworthy that $Ni_{60}Cu_{40}$ in our study is essentially on the trend line in Fig. 4, even though its $T_C$ is somewhat below room temperature (see Fig. 2) where the FMR spin pumping measurements were performed. This result suggests that spin-current dephasing may occur even in the bulk of a metal that is "almost" ferromagnetic with fluctuating magnetic order [41]. Alternatively, the fact that $\lambda_c$ for $Ni_{60}Cu_{40}$ is slightly below the trend line in Fig. 4 may signify that the spin-flip length scale in $Ni_{60}Cu_{40}$ is ≈4 nm, comparable to the dephasing length scale. Though beyond the scope of our present work, the evolution of $\lambda_c$ for Cu content beyond 40 at.% would be an interesting subject for future experiments.

The above-described measurements of $\Delta\alpha$ (Fig. 3) detect spin absorption in the sink, but they provide no direct insight into what the spin current does inside the sink. We therefore examine the byproduct of the transverse spin current interacting with the magnetization: spin-transfer torque. To this end, we employed the synchrotron-based x-ray ferromagnetic resonance (XFMR) technique [24,42–44] at the Advanced Light Source Beamline 4.0.2 [45], which leverages the element-specificity of x-ray magnetic circular dichroism (XMCD). This XFMR technique can directly detect the magnetization dynamics of a specific layer. Moreover, the out-of-plane spin transport here does not involve in-plane net charge transport, hence eliminating ambiguities from coexisting charge-to-spin conversion processes that plague standard electrical spin-torque measurements [46–48].

We conducted XFMR measurements on samples with stack structure MgO(substrate)/Ti(3)/Cu(3)/$Fe_{80}V_{20}$(10)/Ag(5)/Ni(Cu)(5.3)/Ti(3). The (001)-oriented MgO crystal substrate permits high XMCD signals from luminescence yield [45]. As illustrated in Fig. 5(a,b), $Fe_{80}V_{20}$ (instead of $Ni_{80}Fe_{20}$) is the soft low-damping spin source [49,50] for detecting magnetization dynamics via XMCD at the Fe $L_3$ edge – separately from the Ni $L_3$ edge for the Ni(Cu) sink (i.e., Ni or $Ni_{80}Cu_{20}$). The thickness of the Ni(Cu) sink is greater than $\lambda_c$ to ensure complete spin absorption. Our measurements were performed at a microwave excitation frequency of 4 GHz, using a protocol similar to Ref. [51]. We detected the magnetic oscillations transverse to the in-plane applied field by acquiring the XMCD response vs time, as shown in Fig. 5(c,d).

Figure 5(e,f) summarizes the oscillation phase at several values of in-plane applied field $H_x$. The FMR of the FeV source is seen as a 180-degree shift in the phase, $\phi^{src} = \mathrm{atan}(\Delta H/(H_x - H_{FMR}^{src}))$, centered at the resonance field $\mu_0 H_{FMR}^{src} \approx 14$ mT with linewidth $\mu_0 \Delta H \approx 0.95$ mT. For the Ni(Cu) sink, we observe a qualitatively distinct shift in the phase $\phi^{snk}$ around $H_x \approx H_{FMR}^{src}$. We fit $\phi^{snk}$ vs $H_x$ with the following function [42,52],

$$\phi^{snk} - \phi_0^{snk} = \mathrm{atan}\left(\frac{\beta_{dip} \sin^2 \phi^{src} - \beta_{ST} \sin \phi^{src} \cos \phi^{src}}{1 + \beta_{dip} \sin \phi^{src} \cos \phi^{src} + \beta_{ST} \sin^2 \phi^{src}}\right), \quad (2)$$

where $\phi_0^{snk}$ is the baseline phase that depends on the saturation magnetization of the spin sink. The unitless coefficient $\beta_{dip}$ represents the dipolar field torque (e.g., from the interlayer orange-peel coupling [53] with the precessing source magnetization) normalized by the off-resonant microwave field torque. Similarly, $\beta_{ST}$ represents the spin-transfer torque (driven by the pumped spin current [24]) normalized by the off-resonant torque. Since the off-resonant torque scales with the magnetization, $\beta_{ST}$ is also proportional to the efficiency of spin-transfer torque per unit magnetization in the Ni(Cu) sink.

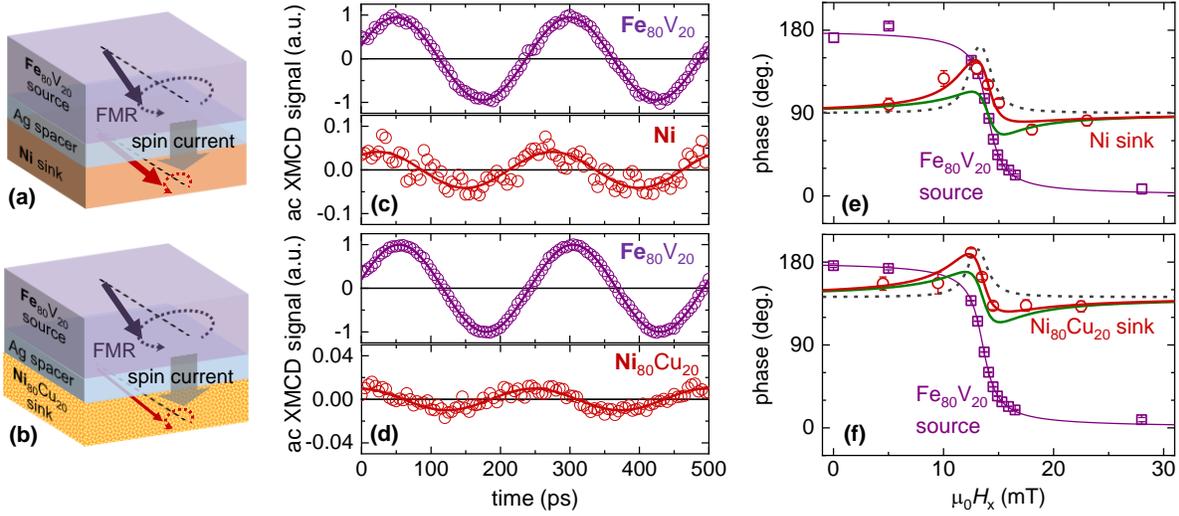

FIG. 5. (a,b) Stack structure for XFMR spin pumping, where the FeV spin source pumps a spin current into the (a) Ni or (b) $Ni_{80}Cu_{20}$ spin sink. (c,d) XMCD response as a function of microwave delay time at the Fe and Ni $L_3$ edges for the samples with the (c) Ni or (d) $Ni_{80}Cu_{20}$ spin sink. The applied field here is $\mu_0 H_x \approx 14$ mT. (e,f) Field ($H_x$) dependence of the oscillation phase for the FeV spin source and the (e) Ni or (f) $Ni_{80}Cu_{20}$ spin sink. The solid red curve represents the fit modeling the total torque in the spin sink; the dashed gray curve represents the contribution from the dipolar field torque (with $\beta_{ST} = 0$ in Eq. 2), and the solid green curve represents the contribution from the spin-transfer torque (with $\beta_{dip} = 0$ in Eq. 2).

The parameters derived from the fitting with Eq. 2 are summarized in Table I. The comparable values of $\beta_{dip}$ for the Ni and $Ni_{80}Cu_{20}$ sinks are reasonable because the dipolar- and microwave-field torques scale similarly with the saturation magnetization of the sink. More importantly, $\beta_{ST}$ also remains the same within experimental uncertainty between Ni and $Ni_{80}Cu_{20}$. We emphasize that $\beta_{ST}$ is an efficiency metric for the spin-transfer torque *per unit magnetization*. Evidently, the Cu impurities do not diminish this spin-transfer torque efficiency. Our finding confirms that a sizable spin-transfer torque emerges from spin dephasing even in an alloy with a high nonmagnetic impurity content. It also implies that spin-transfer torque can be remarkably robust against electronic momentum scattering by impurities.

|  | $\phi_0^{snk}$ (deg.) | $\beta_{dip}$ | $\beta_{ST}$ |
| --- | --- | --- | --- |
| Ni sink | 90 ± 6 | 1.5 ± 0.5 | 1.3 ± 0.5 |
| $Ni_{80}Cu_{20}$ sink | 142 ± 3 | 1.0 ± 0.2 | 1.7 ± 0.3 |

Table I. Parameters for the fit curves of the total torque for the Ni and $Ni_{80}Cu_{20}$ sinks. $\phi_0^{snk}$ is the baseline phase; $\beta_{dip}$ and $\beta_{ST}$ are coefficients proportional to the dipolar field torque and spin-transfer torque, respectively, normalized by the off-resonant microwave field torque

In summary, we have experimentally investigated the mechanism behind the ultrashort coherence length $\lambda_c$ of transverse spin current in ferromagnetic Ni-based thin films. We find that $\lambda_c$ scales inversely with the exchange strength in the ferromagnets examined here, even those with rather high Cu impurity contents. This central result strongly indicates that dephasing – not scattering – dominates transverse spin-current absorption in these nanometer-thick ferromagnetic metals. This result also highlights the ability to tune $\lambda_c$ by engineering the magnetic exchange. While such tuning was previously explored for *ferri*magnets and antiferromagnets [30,54,55], our study demonstrates that $\lambda_c$ can be extended in *ferro*magnets as well by diluting the magnetic order. We further find that the efficiency of spin-transfer torque in a ferromagnet can remain invariant with its impurity content. Our findings provide crucial insights into transverse spin transport in the "bulk" of nanometer-thick ferromagnets, which may help enhance the performance of spin-torque devices by optimizing the length scale of spin dephasing [29].

**Acknowledgments**

Y.L. and S.E. were supported by the Air Force Office of Scientific Research (AFOSR) under Grant No. FA9550-21-1-0365. D.A.S. was supported by the National Science Foundation (NSF) under Grant No. DMR-2003914. This work was made possible by the use of Virginia Tech's Materials Characterization Facility, which is supported by the Institute for Critical Technology and Applied Science, the Macromolecules Innovation Institute, and the Office of the Vice President for Research and Innovation. This research used resources of the Advanced Light Source, a U.S. DOE Office of Science User Facility under Contract No. DE-AC02-05CH11231. S.E. thanks Xin Fan for helpful feedback.

**Data Availability**

The data that support the findings of this study are available from the corresponding author upon reasonable request.